\documentclass[10pt,showpacs,letterpaper,twocolumn,aps,pra]{revtex4-1}
\usepackage[draft]{hyperref}
\usepackage{braket}
\usepackage{amsmath}
\usepackage{graphicx}
\newcommand{\pll}{photoluminescence }

\begin{document}
\preprint{APS/123-QED}
\title{Strong coupling of two quantum emitters to a single light mode:\\ the dissipative Tavis-Cummings ladder}
\author{Nicol\'as Quesada}
\affiliation{McLennan Physical Laboratories, University of Toronto, 60 St. George Street, Toronto, Ontario, Canada M5S 1A7}
\email{nquesada@physics.utoronto.ca}

\begin{abstract} A criterion for strong coupling between two quantum emitters and a single resonant light mode in a cavity is presented.
The criterion takes into account the escape of cavity photons and the spontaneous emission of the emitters, which are modeled as two level systems.
By using such criterion, the dissipative Tavis-Cummings ladder of states is constructed, and it is shown that the inclusion of one more emitter with respect to the Jaynes-Cummings (single emitter) case increases the effective parameter region in which $n^{\text{th}}$ order Rabi splitting is observed. 
\end{abstract}

\pacs{32.70.Jz, 42.50.Ct}

\maketitle
\section{Introduction}
The study of light-matter interaction is one of the most fertile research areas in physics \cite{Haroche06}. 
Under precisely controlled conditions, matter and light can exhibit very interesting phenomenology. 
One such phenomena is the so-called Strong Coupling (SC) regime which is attained in the realm of cavity Quantum Electrodynamics \cite{Walther06}. 
This regime is achieved by isolating the matter and light in such a way that a cavity photon can interact several times with the matter forming an atom-photon ``molecule'' \cite{Carmichael08}. 
Experimentally, the interaction occurs inside a cavity that modifies the spectral density of modes that the atom sees inhibiting spontaneous emission \cite{Kleppner81} and at the same time confining the light \cite{Mabuchi02}.
The formation of this ``dressed'' light-matter state is experimentally verified in the \pll spectrum of the system \cite{Kimble92,Reith04}. 
As the bare matter and light frequencies become resonant, two peaks are observed \cite{Haroche96,Yoshie04}.
This is a manifestation of the formation of dressed states whose frequencies differ under ideal conditions by twice the radiation-matter coupling parameter, the so-called Rabi frequency \cite{Loudon03}. Although this phenomenon is quantum mechanical, the Rabi splitting between the two modes and their widths can be obtained using a simple classical model of two damped harmonic oscillators \cite{Yamamoto07}. 
The criterion for observing the Rabi splitting between one emitter and one mode is given by\cite{Yamamoto07,Tej09,Elena08}:
\begin{equation}\label{lSC}
g>|\gamma_-|,
\end{equation}
where $g$ is the light-matter coupling constant, $\gamma_\pm=(\gamma_a\pm\gamma_\sigma)/4$, $\gamma_a$ is the decay rate of the photons of the cavity, which is inversely proportional to the quality factor of the cavity and $\gamma_\sigma$ is the inverse of the lifetime of the emitter which is related to the Einstein $A$ coefficient \cite{Loudon03}.
The condition given by (\ref{lSC}) implies that the Rabi splitting observed will be given by $R=\sqrt{g^2-\gamma_-^2}$ and that the bare mode (light and matter) populations will oscillate at the same modified Rabi frequency\cite{Tej09}.
Although the Rabi spliting between the modes can be explained using a classical model, the ``quantumness'' of the so-called linear Strong Coupling can be observed by other means, such as the counting statistics of the emitted photons \cite{Yamamoto07,Laussy11,Majumdar11}. 
To evidence quantum behavior in the \pll spectrum not affordable by classical oscillators, one has to consider the case in which more than one photon are bound to the matter. 
In this case, the emission frequencies exhibit \emph{anharmonicity} \cite{Gerry05} and thus, when the system emits a photon different emission peaks will be seen depending on the number of photons bound  \cite{Cirac91,Eberly83}.
The condition for observing Rabi splitting between $n$ photons and one emitter is now given by \cite{Tej09b,Elena08}:
\begin{equation}\label{nlSC}
\sqrt{n}g>|\gamma_-|,
\end{equation}
and the Rabi splitting when there are $n$ photons will be given by the Jaynes-Cummings (JC) modified Rabi frequency:
\begin{equation}\label{JC}
R_n=\sqrt{(\sqrt{n}g)^2-\gamma_-^2}.
\end{equation}
This non linearity has been recently observed for a single emitter being an atom \cite{Schuster08} or a superconducting qubit\cite{Bishop09,Fink08} but remains to be evidenced in systems undergoing strong decoherence such as quantum dots in semiconductor microcavities. 
Much work has been devoted to the study of the emission spectrum of one quantum emitter coupled to a single mode and how decoherence processes such as spontaneous emission, and finite life time of the photon, and different types of incoherent pumping or finite temperature effects affect it \cite{Tej09,Tej09b,Cirac91,Sachdev84,Barnett86,Vera09,Quesada11}. 
The case in which two atoms strongly interact with a single cavity mode has also been studied. 
The evolution of the closed systems was systematically investigated in \cite{Xu92,Luo93,Ami91}. The dynamics of the system in the linear regime was also investigated in \cite{Paulo11,DelValle2} and the evolution of the correlation functions as the system goes from one to several emitters was recently discussed in \cite{Auffeves11}. Finally, experimental results in which the linear regime was explored for the case of two resonant excitons in separated quantum dots and whose Rabi constants are quite similar was presented in \cite{Laucht10}.
Nevertheless, no systematic study has been presented of the dynamics in the non linear regime including decoherence. This is a relevant  problem in quantum dots in microcavitites in which the effects of the environment on the quantum dot are quite strong and more importantly the quality factor of the cavities places the system in an intermediate regime \cite{Tej09b,Laussy11}.
Understanding the dynamics of photons and qubits in a regime where the losses are comparable to the light-matter coupling, is relevant in several areas of quantum information processing in which the light mode acts as an information bus between processor qubits \cite{Pelli95,Ima99,Deutsch03,Simmonds08,Majer07,Blais07} and in general, as a test bed in which multipartite entanglement can be studied \cite{Gywat06,Elena07,Torres10,Restrepo09}. Finally, it is interesting to note that by preparing the initial state of the two emitters in their excited states one has direct access to the non linear regime even in the case where the cavity is empty and this  might be easier to implement than the preparation of 1 or 2 photon states inside the cavity.
In this work then, a criterion for observing strong coupling of two quantum emitters and a light mode in the non-linear regime in which anharmonicities are apparent is presented. 
Although the results derived here are limited to identical quantum emitters where there is no incoherent pumping or finite temperature effects, they serve as the starting point for studying these extensions. \\
The paper has been organized as follows: in section \ref{sec:hamil}, the Hamiltonian dynamics is studied, the eigenenergies and dressed states are written in terms of the coupling constant and the natural frequencies of the emitters and the mode; in section \ref{sec:qrt}, the Master equation that is used to model decoherence in the system is presented and the quantum regression theorem is introduced as the tool used to obtain the dynamics of the first order correlation functions of the subsystems; in section \ref{sec:sc}, by introducing the complex eigenenergies of the system, a criterion for SC is derived in terms of the emission rates of the system and the light matter coupling constant; finally, some conclusions and general remarks are given in section \ref{sec:conc}.

\section{Hamiltonian evolution}\label{sec:hamil}

The system is modeled using the Tavis-Cummings (TC) Hamiltonian \cite{Tavis68} in which each quantum emitter has two possible states, ground $\ket{G}$ and excited $\ket{X}$ and the light is represented by Bosonic annihilation and creation operators ($\hat a, \hat a^{\dagger}$). 
The Hamiltonian, in units in which $\hbar=1$, is given by:
\begin{eqnarray}\label{hamil}
\hat{H}=\omega_0 \hat a^{\dagger}\hat a+\sum_{i=1}^2 \left\{\left(\omega_0-\Delta \right) \hat \sigma_i ^{\dagger} \hat \sigma_i + g (
\hat \sigma_i ^{\dagger} \hat{a} +\hat{a}^{\dagger} \hat \sigma_i   ) \right\},
\label{hamiltonian}
\end{eqnarray}
where $\hat \sigma_i=\ket{G_i}\bra{X_i}$. 
The first two terms of the above equation represent the energies of the photonic mode and the quantum emitters. 
The last term accounts for a dipole interaction between each quantum emitter and the light mode in the Rotating Wave Approximation (RWA). 
The eigenstates of the Hamiltonian for the $n^{\text{th}}$ excitation manifold $\Lambda_n$ in resonance ($\Delta=0$) are given by\cite{Torres10}:
\begin{eqnarray}\label{dressed}
\omega_n^{(1)}&=&\omega_n^{(4)}=\omega_0 n \\
\omega_n^{(2/3)}&=&\omega_0 n \pm  g \sqrt{4n-2}\nonumber\\
\ket{\omega_n^{(1)}}&=&\sqrt {\frac{n}{2n -1}} \ket{n-2,T_1}-\sqrt {\frac {n - 1} {2 n - 1}} \ket{n,T_{-1}} \nonumber\\
\ket{\omega_n^{(2/3)}}&=&\sqrt{\frac{n}{4n-2}} \ket{n,T_{-1}} \pm \frac{1}{\sqrt{2}}\ket{n-1,T_{0}} \nonumber \\
& &+ \sqrt{\frac {n-1} {4 n - 2}}  \ket{n-2,T_{1}}\nonumber\\
\ket{\omega_n^{(4)}}&=&\ket{n-1,S},\nonumber
\end{eqnarray}
where $\ket{n,j} \equiv \ket{n}\ket{j}$ in which the first ket is a Fock state of the field and the second ket is a Dicke state of the matter: $\ket{T_{-1}}=\ket{G_1}\ket{G_2}, \ket{T_{0}}=\frac{1}{\sqrt{2}}\left(\ket{X_1}\ket{G_2}+\ket{G_1}\ket{X_2} \right), \ket{T_1}=\ket{X_1}\ket{X_2}, \ket{S}=\frac{1}{\sqrt{2}}\left(\ket{X_1}\ket{G_2}-\ket{G_1}\ket{X_2} \right)$.
The dressed state and energy for the lowest excitation manifold is simply $\ket{0,T_{-1}}$ with zero energy. 
For $n=1$ the dressed states and energies are $\ket{\omega_1^{(2)}}, \ket{\omega_1^{(3)}}, \ket{\omega_1^{(4)}}$ with energies $\omega_1^{(2)}, \omega_1^{(3)}, \omega_1^{(4)}$ given by (\ref{dressed}). 
Whenever the losses are very small, the emission spectrum of the system will consist of transitions in which the system goes from a dressed state with $n$ excitations to another with $n-1$ excitations, and the positions of the peaks in the \pll spectrum  will be given by the difference between their energies, 
\begin{equation}\label{emission}
\tilde \nu^{n \to n-1}=\omega_{n}^{(i)}-\omega_{n-1}^{(j)}. 
\end{equation}
Finally, note if the emitters where not identical then the eigenenergies for the $n^{\text{th}}$ excitation manifold would be the roots of a quartic polynomial whose roots could only be expressed in terms of Ferraris solution\cite{Abra72} which are much complicated than the ones presented here and are of very limited use.

\section{Dissipative dynamics}\label{sec:qrt}
\begin{figure}
\includegraphics[width=0.48\textwidth]{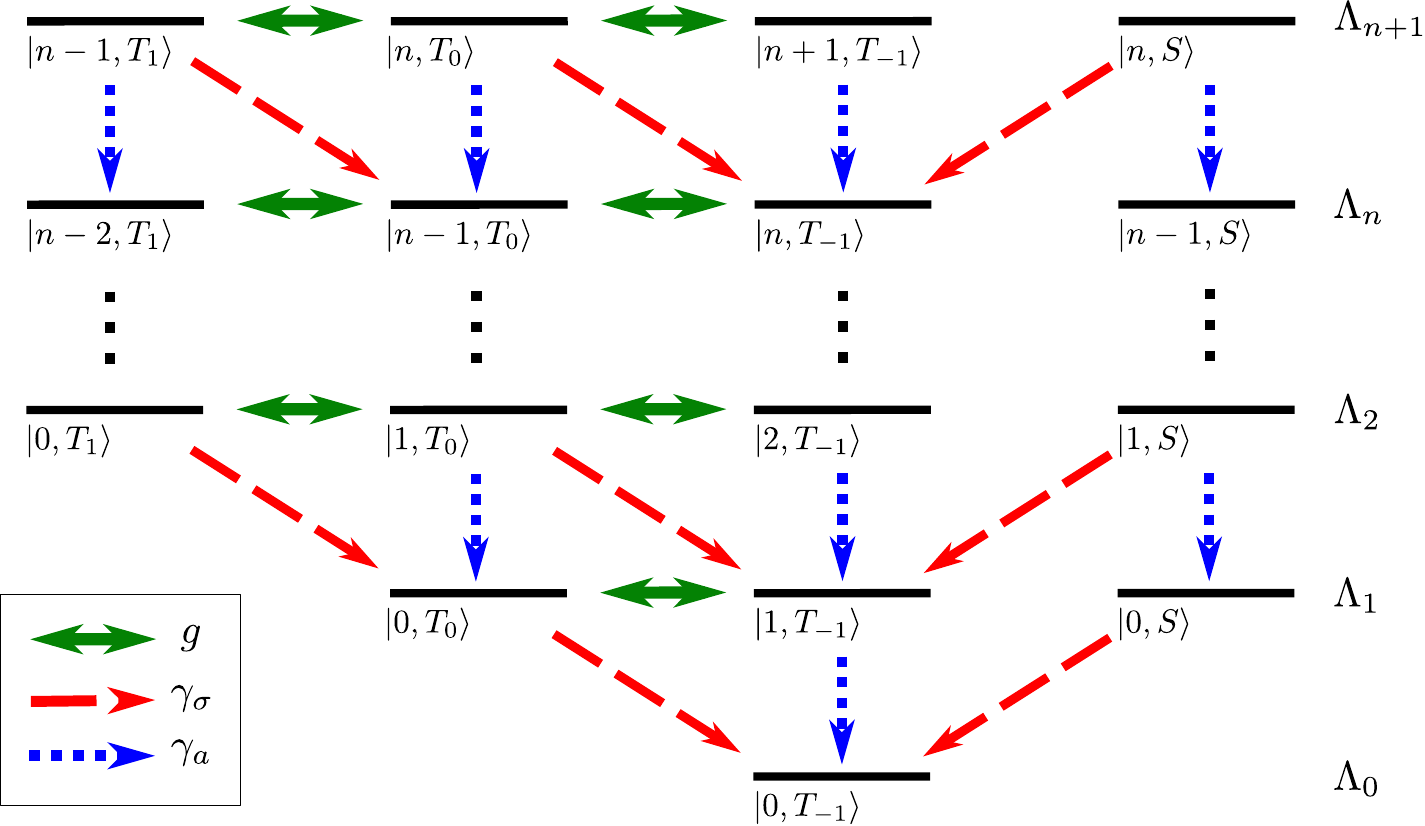}
\caption{\label{ladder} (Color online) Ladder of bare states at resonance ($\Delta=0$) for two 2-level systems coupled to a cavity mode for the first three excitation manifolds and for two excitation manifolds containing $n$ and $n+1$ excitations. 
The double headed solid green horizontal arrows depict the light-matter coupling $g$, dashed red diagonal arrows the spontaneous emission process $\gamma_\sigma$ and blue dotted vertical arrows the coherent emission process $\gamma_a$. 
It is interesting to note that, because the spontaneous emission process acts independently on each two-level system, it is the only process that couples the singlet state $\ket{S}$ to the dynamics. The spacing between each excitation manifold $\Lambda_n$ is given by $\omega_0$. In the non resonant case the atomic states with one excitation $\ket{n,0}, \ket{n,S}$ will be shifted downwards by an amount $\Delta$ and the state with zero atomic excitations $\ket{n,-1}$ will be shifted by an amount $2 \Delta$.}
\end{figure}

To fully account for the escape of photons outside the cavity and the spontaneous emission of the emitters, the dissipative dynamics of the system must be studied.
This can be done by writing a Lindblad master equation for the density operator of the system that accounts
for coherent emission of photons (with rate $\gamma_a$) and spontaneous emission
of the emitters (with rate $\gamma_\sigma$). Such master equation is given by \cite{DelValle2}:
\begin{eqnarray}\label{eq:master}
 \frac{d}{dt} \hat \rho = i [\hat \rho, \hat H]+ \frac{\gamma_a}{2}
\mathcal{L}_{\hat a}\{\hat \rho\}+\frac{\gamma_\sigma}{2} \sum_{i=1}^2 \mathcal{L}_{\hat \sigma_i} \{ \hat \rho \}
\end{eqnarray}
where $\mathcal{L}_{\hat O}\{\hat \rho\}=2 \hat O \hat \rho \hat O^{\dagger}-\hat O^{\dagger} \hat O \hat \rho-\hat \rho \hat O^{\dagger} \hat O$. A depiction of the processes included in the above master equation in the ladder of bare states is given in figure \ref{ladder}.

To obtain the emission spectrum of the field or the emitters, it is necessary to write the first order correlation:
\begin{equation}\label{corr}
\mathcal{G}_{\hat O} (t,\tau)=\braket{\hat O^{\dagger}(t+\tau)\hat O(t)},
\end{equation}
where $\hat O$ can be any of $\hat \sigma_1, \hat \sigma_2$, $\hat a $. Once $\mathcal{G}_{\hat O} (t,\tau)$ is at hand, the emission spectrum is given by\cite{Eberly77}:
\begin{eqnarray}\label{fourier}
& &\mathcal{S}_{\hat O}(\omega,T)=2 \kappa \times\\
& &  \Re \left\{\int_0^T d \tau e^{(\kappa-i\{\omega-\omega_0\}) \tau} \int_0^{T-\tau} dt e^{-2 \kappa (T-t)} \mathcal{G}_{\hat O}(t,\tau) \right\} \nonumber
\end{eqnarray}
where $\kappa$ is the finite bandwidth of the spectrometer and $T$ is the amount of time in which light has been collected.
To obtain the delayed time $\tau$ dynamics of the correlation functions, the Quantum Regression Theorem (QRT) is used \cite{Walls94}. It asserts that, given a set of operators $\hat X_j$ satisfying the single time $t$ dynamics,\\
\begin{equation}\label{premiseQRT}
\frac{\partial }{\partial t} \braket{\hat X_j(t)} = -i \sum_k L_{jk} \braket{\hat X_k(t)},
\end{equation}
then the two-time dynamics with an arbitrary operator $\hat Y$ is given by:
\begin{equation}
\frac{\partial }{\partial \tau} \braket{\hat X_j(t+\tau) \hat Y(t)} = -i \sum_k L_{jk} \braket{\hat X_k(t+\tau)\hat Y(t)},  
\end{equation}
for any operator $\hat Y$. 
It can be easily seen that $\hat a^{\dagger}, \sigma_1^{\dagger}$ and $\sigma_2^{\dagger}$ can be written as linear combination of the basis set, $\hat A_{l,j;m,i}^{\dagger}=\ket{l,j}\bra{m,i}$  with $\ket{l,j}\in \Lambda_{n-1}$ and $\ket{m,i}\in \Lambda_{n-1}$ and that because the TC hamiltonian preserves the number of excitations $\hat N = \hat a^{\dagger} \hat a+ \sum_{i=1}^2\hat \sigma_i^{\dagger} \hat \sigma_i$ they will satisfy the premise of the QRT.
The expected values of the operators $\braket{\hat A^\dagger_{l,j;m,i}(t)}$ can then be arranged in a vector, $\mathbf{x}(t)$ that will satisfy (\ref{premiseQRT}) which can be written as: $\frac{\partial}{\partial t} \mathbf{x}(t) =-i \mathbf{L}  \mathbf{x}(t)$ where $\mathbf{L}$ is a square matrix. 
If the two time expected values $\braket{\hat A^\dagger_{l,j;m,i}(t+\tau)w(t)}$ ($w \in \{ \hat \sigma_1, \hat \sigma_2, \hat a \}$) are also arranged in a vector $\mathbf{v}(t+\tau,t)$, then, they will satisfy the same differential equation with respect to $\tau$, $\frac{\partial}{\partial \tau} \mathbf{v}(t+\tau,t) =-i \mathbf{L} \mathbf{v}(t+\tau,t)$.\\ 
By ordering the operators $\hat A^\dagger_{l,j;m,i}$ according to the excitation manifolds they connect, the matrix $\mathbf{L}$ takes a block upper triangular form whose eigenvalues define the widths ($\tilde \Gamma_k$) and positions ($\tilde \nu_k$) of the emission peaks by $\tilde \lambda_k= -i\tilde \Gamma_k+ \tilde \nu_k$. \\
Note that, to use the QRT, the $\tau=0$ initial conditions are required; these are given by $\braket{\hat X_k(t+\tau)\hat Y(t)}|_{\tau=0}=\braket{\hat X_k(t)\hat Y(t)}=\text{tr}(\hat \rho(t) \hat X_k \hat Y)$. The $t$ dynamics of the initial conditions required for the QRT can be studied in a similar fashion to the $\tau$ dynamics.
To this end, one sets $\tau=0$ to obtain the dynamics of the populations of each subsystem, $\mathcal{G}_{\hat O}(t,\tau=0)=\braket{\hat O^{\dagger}(t)\hat O(t)} =\text{tr}(\hat \rho(t) \hat O^{\dagger} \hat O)$.
In this case, it is easily seen that the required operators are linear combinations of the set $\hat F_{l,j;m,i}=\ket{l,j}\bra{m,i}$ with $\ket{l,j},\ket{m,i} \in \Lambda_n$ and that this set also makes a closed set of differential equations.
The expectation values of such operators can be organized in another vector $\mathbf{y}(t)$ that satisfies a differential equation of the form $\frac{\partial}{\partial t} \mathbf{y}(t)= -i\mathbf{D} \mathbf{y}(t)$ where the matrix $\mathbf{D}$ is block upper triangular.

\begin{figure}
\includegraphics[width=0.48\textwidth]{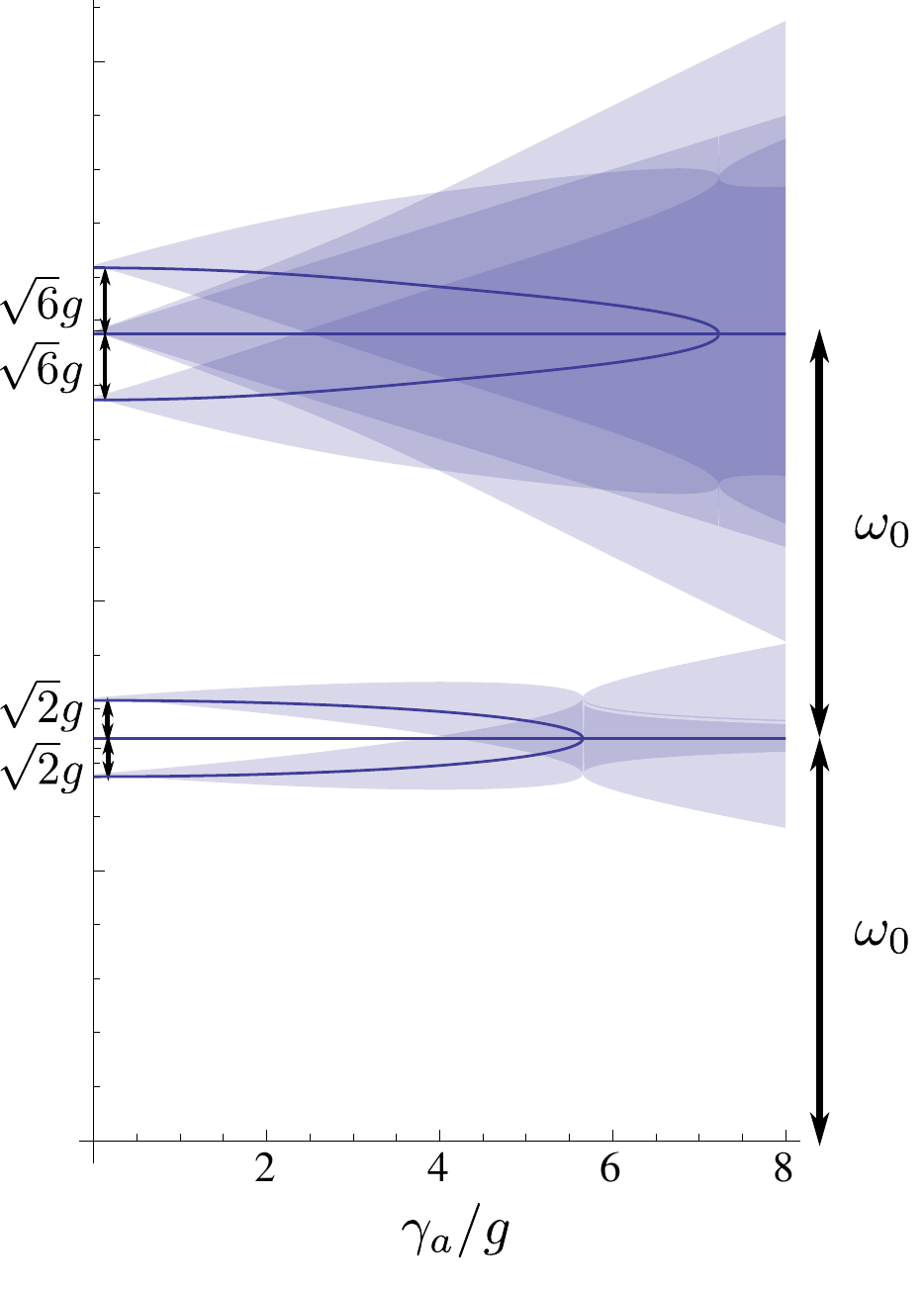}
\caption{\label{cenergiesplot} (Color online) Positions (thick lines) and widths (shadowing) of the complex eigenenergies at resonance for the first two rungs of the system as a function of the inverse life-time of the photon $\gamma_a$. The spontaneous decay rate of the system has been set to zero ($\gamma_\sigma=0$). Note that in experiments typically $\omega_0 \gg g$ and thus the widths between different excitation manifolds do not overlap. Also note that for more than one excitation there are always two eigenenergies that are degenerate but that differ in their associated width which is illustrated here for the second rung of the ladder.
}
\end{figure}

\section{Strong coupling criterion and the complex eigenenergies}\label{sec:sc}
\begin{figure}
\centering
\includegraphics[width=0.45\textwidth]{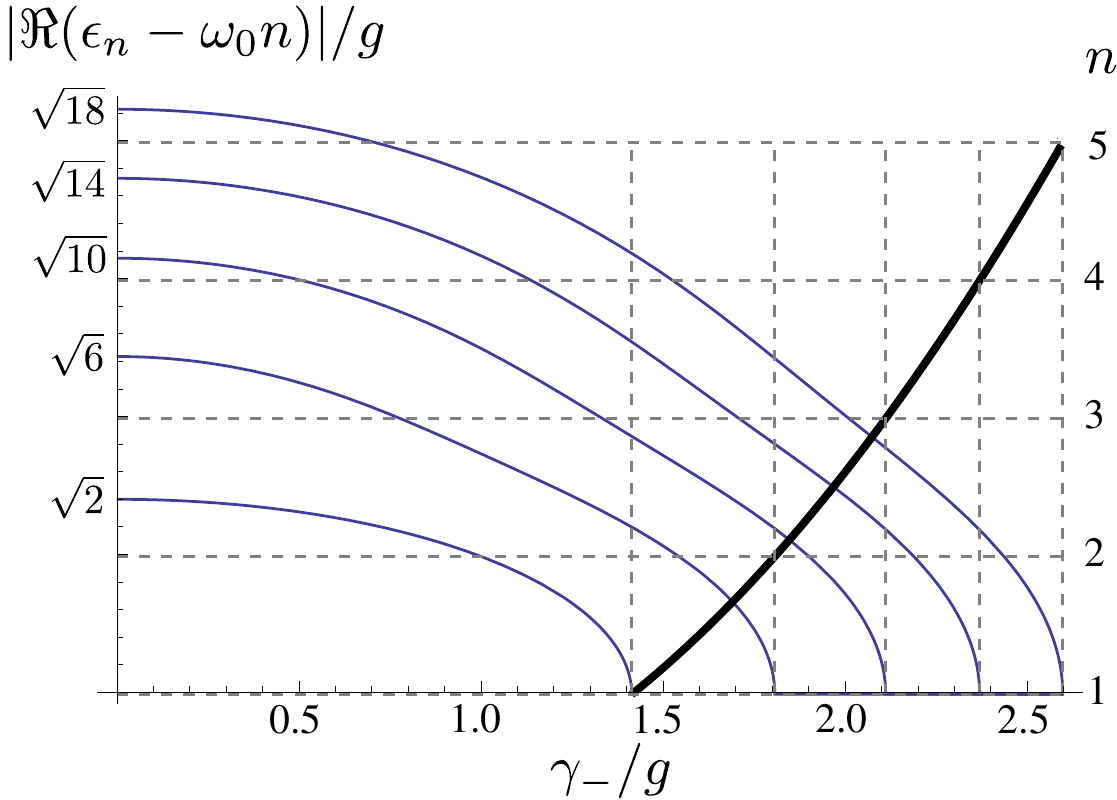}
\caption{\label{critt}(Color online) The black thick line is the contour of $|\Im(\mathcal{Q}_n)|=1$ at resonance $\Delta=0$ , $n$ is taken as a real number whose scale is in the right of the figure. The purple thin lines are the different Rabi splittings $|\Re(\epsilon_n-\omega_0 n)|$  at resonance $\Delta=0$ for the first 4 excitation manifolds (the scale of the different splittings is in the left of the figure). Note that whenever the contour of $|\Im(\mathcal{Q}_n)|=1$ crosses an integer value of $n$ as $\gamma_-/g$ is varied one of the Rabi splittings becomes zero. Finally, as expected from the purely Hamiltonian picture the $n^{\text{th}}$ order Rabi splitting reduces to $\sqrt{4n-2}$ as the dissipation effects become negligible $\gamma_-/g \to 0$.}
\end{figure}

In this section, the eigenvalues of the regression matrix $\mathbf{L}$ and the population matrix $\mathbf{D}$ are obtained and based on their dependence on the system parameters a criterion for observing Rabi splitting between the different transitions is derived. 
It is easily seen that $\mathbf{L}= \omega_0 \mathbf{1}+\mathbf{B}$ where $\mathbf{B}$ depends on $g,\gamma_a,\gamma_\sigma$ and $\Delta$ but not $\omega_0$, and $\mathbf{1}$ is the identity matrix. 
The Rabi splitting will then be  determined by the properties of $\mathbf{L}$.
In a similar way one can note that the population matrix $\mathbf{D}$ is also independent of $\omega_0$, and that the eigenvalues of both matrices can be written in terms of the complex eigenenergies in straightforward generalization of equation (\ref{emission}).
To this end in this section the complex eigenenergies of the Liouvillian (\ref{eq:master}) are introduced. In this case the real part will contain the information of the resonances of the system and the imaginary part will give information about its width. As we shall see the eigenenergies will still contain the symmetries of the original Hamiltonian, \emph{i.e.} they will come in a triplet and singlet that correspond to those in equation (\ref{dressed}) but now will have a non zero imaginary part that will account for the dissipation. Also as we shall show as the dissipation grows large the real part of the eigenenergies will be given only by $n \omega_0-\Delta\pm \Delta$ and any trace of the coupling $g$ will disappear.
Note that even in the dissipative case the complex eigenenergy of the lowest energy-state, the light matter vacuum $\ket{0,T_{-1}}$, is still strictly zero and is not affected by the dissipation. This is merely a reflection of the fact that the environment that was traced to obtain the master equation (\ref{eq:master}) is a zero temperature reservoir and thus the system will tend in the long time to reach its lowest energy eigenstate. For the first excitation manifold the complex eigenenergies are simply given by:
\begin{eqnarray}
 \epsilon_1^{(1/2)}&=&-i \gamma_+ +  \omega_0-\Delta \pm  \mathcal{R}_1\\
 \epsilon_1^{3}&=&-i\frac{\gamma_{\sigma}}{2}+ \omega_0-\Delta . \nonumber 
\end{eqnarray}
where the first order complex Rabi frequency is given by:
\begin{eqnarray}
\mathcal{R}_1=\sqrt{2 g^2-\left(\gamma_- +i \frac{\Delta}{2}\right)^2}
\end{eqnarray}
These are precisely the same energies that are obtained in the linear regime \cite{Paulo11,DelValle2}, by treating the operators $\hat \sigma_i$ as bosons, and naturally reduce  to the purely real eigenenergies of the Hamiltonian as $\gamma_a$ and $\gamma_\sigma$ go to zero.
From demanding that the modified Rabi frequency $\mathcal{R}_1$ be real at zero detuning  the linear strong coupling condition is derived:
\begin{eqnarray}
\sqrt{2}g > |\gamma_-|
\end{eqnarray}
 At zero detuning in the SC regime the widths of $\epsilon_1^{(1/2)}$ are equal which stems from the fact that in the SC regime the dressed state picture is to some extent  valid and because of this at $\Delta=0$ the two dressed states are half matter-half light and thus decay with an average of the two decay rates. Also note that the decay rate or width associated with $\epsilon_1^{(3)}$ is purely due to $\gamma_{\sigma}$ which also follows from the fact that the purely matter state $\ket{0,T_0}$ is an eigenstate of $\hat H$.\\
For more than one excitation ($n>0$) the complex eigenenergies of the problem are simply given:
\begin{eqnarray}\label{cenergies}
\epsilon_n^{(1,2,3)}&=& -i\frac{\Gamma_n}{2} +  n \omega_0 -\Delta +  \mathcal{P}_n^{(1,2,3)}  \\
  \epsilon_n^{(4)}&=& -i\frac{\Gamma_n}{2} +  n \omega_0-\Delta, \nonumber 
\end{eqnarray}
where the widths $\Gamma_n$ are given by:
\begin{equation}
\Gamma_n=(n-1)\gamma_a+\gamma_\sigma 
\end{equation}
and,
\begin{equation}\label{pn}
\mathcal{P}_n^{(k)}= \mathcal{R}_n \cos \left(\frac{ \cos^{-1}\left[ i \mathcal{Q}_n   \right] + 2 k \pi }{3}\right)/\cos\left(\pi/6 \right),   \\
\end{equation}
is given in terms of the discriminant \footnote{$\mathcal{P}_n^{(k)}$ are the roots of the cubic polynomial in $x$: $x^3+x \left((4 n-2) g^2-\left(2 \gamma _-+i \Delta \right){}^2\right)-2\left(2 \gamma _- + i \Delta \right)  g^2  $
}
\begin{equation}\label{qn}
\mathcal{Q}_n=\frac{6\sqrt{3} \left(\gamma_-+i\frac{\Delta}{2}\right)/g}{\left(\mathcal{R}_n/g \right)^3} \\
\end{equation}
and the complex Rabi frequency:
\begin{equation}\label{rn}
 \mathcal{R}_n=\sqrt{(4n-2)g^2-4\left( \gamma_-+i\frac{ \Delta}{2}\right)^2}.
\end{equation}
The complex eigenenergies are plotted in Figure (\ref{cenergiesplot}) as a function of the inverse lifetime of the photon $\gamma_a$.
As it was mentioned before the eigenenergies still retain some of the symmetries of the Hamiltonian, \emph{i.e.} they are splitted between a triplet and a singlet.
Because the singlet $\ket{n-1,S}$ whose energy is $\epsilon_n^{(4)}$ does not couple coherently to the other states of the system it simply acquires a non-zero imaginary part that corresponds to $(n-1)$ times the decay of a single photon plus the decay rate of a single atom.\\ Far more interesting is what happens to the triplet of states. First note that the discriminant $\mathcal{Q}_n$ is a function only of the quantity $\gamma_-+i \Delta/2$. This simply points to the fact that as far as the splitting between the different energies in a given excitation manifold is concerned, the decay rates act only as an ``imaginary'' detuning. \\
To define the transition between strong and weak coupling it is necessary to study how the Rabi splitting at zero detuning ($\Delta=0$) between different energies in a given excitation manifold changes and in particular when does it become zero. 
Naively, it might be thought that the necessary and sufficient condition to have complex roots with non-zero imaginary parts is that the modified Rabi frequency, $\mathcal{R}_n$ be real (and thus automatically $\mathcal{Q}_n$ will be as well); this condition would be given by:
\begin{equation}\label{rabi1}
\sqrt{4n-2}g>2 |\gamma_-|.
\end{equation}
Nevertheless, the condition given by the above equation is overly restrictive. 
Even in the case in which $\mathcal{R}_n$ is purely imaginary, the splitting given by (\ref{pn}) can have non zero real parts.
Note that, when $\mathcal{R}_n=i \Im(\mathcal{R}_n)$ is purely imaginary, then also $\mathcal{Q}_n=i \Im(\mathcal{Q}_n)$ and thus, the argument of the $\cos^{-1}$ in (\ref{qn}) is purely real; nevertheless, the $\cos^{-1}$ is a real number if its argument is real \emph{and} in absolute value less than or equal to one. Thus even when $\mathcal{R}_n$ is purely imaginary  if  $|\Im(\mathcal{Q}_n)|\geq 1$ there will be $n^{\text{th}}$ order Rabi splitting.
\emph{In summary, the most general condition for having non zero Rabi splitting in the $n^{\text{th}}$ excitation manifold is, assuming that the complex Rabi frequency at zero detuning is purely imaginary, given explicitly by}:
\begin{equation}\label{crite}
|\Im(\mathcal{Q}_n)|= \frac{6 \sqrt{3}|\gamma_-/g|}{\left(\sqrt{\left(2\gamma_-/g\right)^2-\left(4n-2\right)}\right)^3}>1.
\end{equation}
Again, note that if the Rabi frequency $\mathcal{R}_n$ is real at zero detuning then automatically the system is in SC. Only for cases when it becomes imaginary the above criterion is needed. In figure \ref{critt} the contour of $|\Im(\mathcal{Q}_n)|=1$ is plotted as a function of $n$ and $\gamma_-/g$. It is clearly seen that whenever the contour crosses an integer $n$ the Rabi splitting of the $n^{\text{th}}$ rung becomes zero.\\
In the limit $\gamma_-/g \ll 1$, one can expand (\ref{pn}) to second order in $\gamma_-/g$:
\begin{eqnarray}\label{approx}
 \mathcal{P}_n^{(k)}  \approx \left\{ 
  \begin{array}{l}
     g \sqrt{4 n-2} +i\frac{\gamma _-}{2n-1} -  g \frac{ \left(16 n(n-1)+1\right) }{2^{3/2} (2 n-1)^{5/2}}\left(\frac{\gamma _-}{g} \right)^2  \\
        -i \frac{2 \gamma _-}{2 n-1}  \\
     -  g \sqrt{4 n-2} +i\frac{\gamma _-}{2n-1} +  g \frac{ \left(16 n(n-1)+1\right) }{2^{3/2} (2 n-1)^{5/2}}\left(\frac{\gamma _-}{g} \right)^2  \\   
  \end{array} \right.
\end{eqnarray}
which explicitly shows that as $\gamma_a$ and $\gamma_{\sigma}$ go to zero the complex eigenenergies become purely real and reduce to the eigenenergies of the Hamiltonian (\ref{hamil}).\\
Going back to the dynamics of the first order correlation function $\mathcal{G}_{\hat O}(t,\tau)$, the matrix that characterizes its dynamics, $\mathbf{L}$ will be block upper triangular. Each block will represent the emission of one photon after the system decays from one excitation manifold to the one that is immediately below it. The matrix then will have a 3 dimensional block representing the transitions between the first excitation manifold and the vacuum, then a 12 dimensional block representing the transitions between the second and first excitations (with 4 and 3 eigenenergies that give 12 possible transition frequencies) and finally 16 dimensional blocks representing transitions between the 4 possible energies of two contiguous excitation manifolds. The eigenvalues corresponding to the $m^{\text{th}}$ block will be given by 
\begin{eqnarray}\label{diff}
\lambda^{i,j}_m=\epsilon_m^{i}-\left(\epsilon_{m-1}^{j}\right)^*,
\end{eqnarray}
where $z^*$ is the complex conjugate of $z$.
The last equation tells that to obtain the transition frequencies the real parts are subtracted (in the Hamiltonian case this is precisely what is done (\ref{emission})) and the imaginary parts are added to obtain the width of the emission line.
Finally, it is worth mentioning that as can be seen from figure (\ref{cenergiesplot}) at $\Delta=0$ a pair of energies become degenerate and thus fewer peaks will be seen. For transitions between two excitation manifolds other than the one involving the vacuum, this will imply only nine transitions will give photons with different energies, and thus at most nine emission peaks could be observed. This number is further reduced if the initial condition of the system is such that its density operator has projections only in the symmetric subspace spanned by the triplet states $\ket{n,T_i}$. In this case the system will not be able to transit through the singlet state $\ket{n,S}$ and thus the contribution associated with the energy $\epsilon_n^{(4)}$ will not appear in equation (\ref{diff}).
As for the dynamics of the occupation numbers which serve as the initial conditions for the QRT the matrix $\mathbf{D}$ that characterizes its dynamics will also be block diagonal but know the blocks will correspond to energy differences between the same excitation manifold: 
\begin{eqnarray}
\delta_m^{i,j}=\epsilon_m^{i}-\left(\epsilon_{m}^{j}\right)^*.
\end{eqnarray}
 Note that for the excitation manifold that corresponds to the vacuum there is only one energy $\epsilon_0=0$ and only one $\delta_0=0$ which simply accounts for the conservation of probability ($\text{tr} (\hat \rho(t))=1$) that the master equation (\ref{eq:master}) provides.

\section{Conclusions}\label{sec:conc}
A criterion for observing Rabi splitting between two quantum emitters and a single cavity mode has been presented in terms of the Rabi splitting of the complex eigenenergies of the master equation (\ref{eq:master}). It has been shown that the criterion given by (\ref{crite}) is a robust characteristic of the dynamics given by (\ref{eq:master}) since it accounts both for the delayed dynamics, which is relevant for the calculation of the first order correlation function and also the dynamics of the density matrix populations which contains the information about the occupation numbers of the subsystems.
The precise positions of the emission peaks were given in terms of the real parts of the complex eigenenergies and the rich multiplet structure of the \pll spectrum was discussed. 
An intuitive picture of the possible types of multiplets was given in terms of whether each rung of the Tavis-Cummings ladder exhibited Rabi splitting or not.
Although the condition given by (\ref{crite}) determines the presence of oscillatory frequencies in the dynamics of the first order correlation function that will lead to anti-crossings in the \pll spectrum, this anti-crossings will not necessarily be easily resolved as can be seen in figure \ref{cenergiesplot}. 
The reason for this is that the broadening of the spectral lines, which are given by the imaginary parts of the eigenenergies (\ref{cenergies}), grows at least linearly in $n \gamma_a$. 
On the other hand,the spacing between the lines grows approximately as $\sqrt{n}g$ as can be seen from (\ref{approx}). 
Because of this last observation, it is more feasible to observe the anharmonicities described here by focusing on suppressing the emission of the cavity mode, \emph{i.e.}, in very high $Q$ cavities. 
The results presented here allow to analyze the differences between the well known single emitter (JC) case in the dissipative regime and the case were two emitters are present (TC).
It was already well known that the number of emission lines increases significantly; this is simply because each excitation manifold will have more states and thus more transitions can occur. 
With the results presented here, also the dissipative open system dynamics can be compared. 
For instance, in the JC model, all the dynamics is determined by whether the modified Rabi frequency (\ref{JC}) of the one emitter case is real or not, whereas in the case of two emitters, a more involved criterion is necessary (\ref{crite}). The system with two emitters is more robust against decoherence since the parameter region in which SC can be observed is bigger as compared to the area in which the two emitter Rabi frequency is real. 
It is also interesting to note that, in the JC case, $\gamma_-$ only shrinks the Rabi splitting between the lines but does not modify the broadening of the spectral whereas, in the two emitter case, it does, as can be seen from (\ref{approx}). 
Thus, this work gives insights in the interplay between cooperativity and dissipation by analyzing the simplest case of cooperative effects in the interaction between light and matter under decoherence effects.

\begin{acknowledgments}
The author is grateful to B. A. Rodr\'iguez, H. Vinck-Posada and P.C. C\'ardenas for enlightening discussions, also thanks D.F.V. James for a critical reading of the manuscript and for providing many valuable comments and acknowledges financial support from a University of Toronto fellowship.
\end{acknowledgments}

\end{document}